\documentclass[twocolumn,showpacs,preprintnumbers,amsmath,amssymb,prl]{revtex4}
\usepackage[dvips]{graphicx}% Include figure files
\usepackage{dcolumn}% Align table columns on decimal point
\usepackage{bm}% bold math

\begin{document}

%\preprint{APS/123-QED}

\title{Single-Particle Spectrum in the Electron-Doped Cuprates}

\author{Hiroaki Kusunose}%\email{hiro@itp.phys.ethz.ch}
\author{T.M. Rice}
\affiliation{Institut f\"ur Theoretische Physik, ETH-H\"onggerberg,
  CH-8093 Z\"urich, Switzerland}

\date{\today}

\begin{abstract}
We study the evolution of the single-particle spectrum with electron doping in a scheme which adds multiple exchange of transverse spin excitations to the mean-field antiferromagnetic insulator.
Away from half-filling small Fermi surface pockets appear first around the X-points, and simultaneously new spectral weight grows in the insulating gap.
With further doping the in-gap states develop the character of a renormalized quasiparticle band near the chemical potential.
The essential features in momentum-energy space agree well with recent studies using angle-resolved photoemission spectroscopy on electron-doped cuprates.
We interpret the origins and the nature of the in-gap states using a simple variational wavefunction.
\end{abstract}

\maketitle

%.Introduction
Recent angle-resolved photoemission spectroscopy (ARPES) experiments have determined the evolution of the Fermi surface in electron-doped cuprates, Nd$_{2-x}$Ce$_x$CuO$_{4\pm\delta}$ \cite{Armitage02}.
In contrast to the hole-doped case, small Fermi surface pockets appear first near the X-points and then with further doping still in the presence of antiferromagnetic (AF) order, in-gap states appear around the zone diagonals with the character of renormalized band pockets.
A further increase in doping leads to the formation of a single large Fermi surface.
Note that ARPES for electron doping is more informative than for hole doping since the evolution of the charge-transfer gap in the single-particle spectrum can be directly observed.
In this Letter we present a systematic investigation of the single-particle properties of electron-doped antiferromagnets using parameters appropriate for the cuprates, and discuss the physical origins of the in-gap states.

The ARPES experiments give clear evidence for electron-hole asymmetry, in agreement with the significantly different behavior of electron- and hole-doped cuprates \cite{Armitage02,Damascelli03}.
Similar contrasting behavior was found in recent one-loop renormalization-group studies \cite{Honerkamp01}. A clear separation in energy scale between AF and superconducting fluctuations appears for electron-doping \cite{Honerkamp01,Manske00,Kobayashi02}.
This explains the robustness of the AF phase \cite{Chubukov95} and the small doping range of the superconducting phase.
For this reason, electron-doped cuprates can be treated as doped antiferromagnets within a coupling weak to moderate scheme, in contrast to hole-doped cuprates, where the AF fluctuations compete with the superconducting fluctuations leading to a crossover at relatively high energy scale to a strong coupling phase where all play a role \cite{Honerkamp01,Furukawa98,Honerkamp01a,Yanase01,Kobayashi01}.
Strong coupling approaches give strong asymmetry \cite{Lee03} but the narrow region of the AF order upon hole doping greatly limits the approach we use here which starts from the AF mean-field (MF) state.

Since the AF Mott insulating state may be adiabatically connected to the MF-AF state at $T=0$, the MF-AF state can be a good starting point for our purpose \cite{Kusko02,Schrieffer89,Bulut94}.
In order to take account of the collective spin-wave excitations (and also the particle-hole continuum), we calculate the single-particle self-energy including multiple exchange of RPA-type transverse spin excitations, {\it i.e.}, so-called self-consistent Born approximation (SCBA) \cite{Altmann95,Brenig95}.
Within this approach we can demonstrate the following. Away from half-filling small Fermi surface pockets occur first around the X-points, and simultaneously new spectral weight appears in the insulating gap.
With further doping the in-gap states acquire the features of a renormalized quasiparticle band showing clear resemblance to that in the paramagnetic phase.
To explain the physical origin of the in-gap state we introduce a simple variational wavefunction for the in-gap state, in which the quasihole MF state in the lower band admixes strongly with the hole state accompanied by particle-hole spin excitations in the upper band.

We work with the $t$-$t'$-$U$ Hubbard model,
\begin{equation}
H=\sum_{{\bm k}\sigma}\epsilon_{{\bm k}}c^\dagger_{{\bm k}\sigma}c_{{\bm k}\sigma}+U\sum_in_{i\uparrow}n_{i\downarrow},
\end{equation}
with a next-nearest-neighbor tight-binding dispersion $\epsilon_{\bm{k}}=-2t(\cos k_x+\cos k_y)-4t'\cos k_x\cos k_y$, where the lattice spacing $a=1$.
Throughout this paper, we set $t'/t=-0.3$ and $U/t=8$.
The outline of the SCBA calculation is as follows.
First, we solve the AF-MF equation to determine the MF gap $\Delta_0$ and the MF chemical potential $\mu_0$ for a given electron density $n$:
\begin{eqnarray}
&&\frac{1}{U}=-\frac{1}{N}\sum_{\bm k}^{\rm MBZ}\sum_m^\pm\sum_\sigma^{\uparrow,\downarrow} \theta(\mu_0-E_{{\bm k}m}) \frac{m}{2E^0_{\bm k}}, \\
&&n=\frac{1}{N}\sum_{\bm k}^{\rm MBZ}\sum_m^\pm\sum_\sigma^{\uparrow,\downarrow}\theta(\mu_0-E_{{\bm k}m}),
\end{eqnarray}
where the MF-quasiparticle energy in the upper (lower) band, $m=+1(-1)$, is given by $E_{{\bm k}m}=\xi^+_{\bm k}+mE^0_{\bm k}$, where $\xi_{\bm k}^\pm=(\epsilon_{{\bm k}+{\bm Q}}\pm\epsilon_{\bm k})/2$ and $E^0_{\bm k}=\sqrt{(\xi^-_{\bm k})^2+\Delta_0^2}$.
Here $N$ is the number of sites and the summation of ${\bm k}$ is taken over the magnetic Brillouin zone (MBZ).
Note that the AF long-range order is always present in the doping range we study in this paper.
We set up the RPA transverse susceptibilities (for normal and umklapp processes) using $\Delta_0$ and $\mu_0$, and solve the Dyson equation for the retarded Green's functions at $T=0$ putting these RPA susceptibilities into the one-loop self-energy.
The true chemical potential, $\mu$, is determined by $n=2\int_{-\infty}^\mu d\omega\rho(\omega)$, where $\rho(\omega)$ is the single-particle density of states (DOS) calculated from the resultant Green's function.
We discretize the momentum by $32\times32$ mesh in the first quadrant and the energy by $1024$ mesh in the interval $(-20t,20t)$.
Due to the omission of vertex corrections, the coupling constant in the self energy is too strong, which overestimates renormalization of the indirect gap \cite{Altmann95,Brenig95}.
As shown by Bulut {\it et al.} \cite{Bulut93}, a reduction of the coupling constant $U$ mimics the effect of vertex corrections.
We obtain a reasonable value of the indirect gap at half-filling with the reduced coupling constant, $0.7U$, in the self energy.

\begin{figure}[t]
\includegraphics[width=8.5cm]{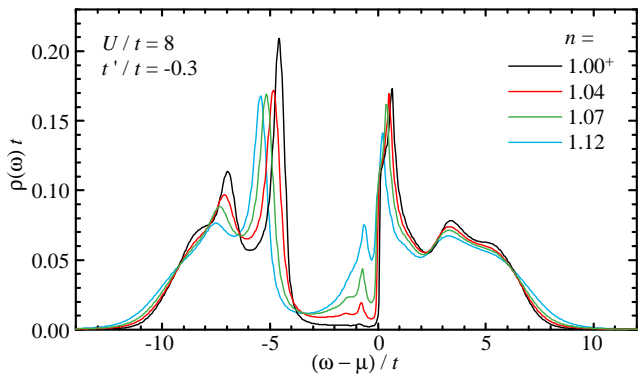}
\caption{The doping dependence of the DOS. Sharp quasiparticle peaks appear at the edges of the insulating gap with broad incoherent backgrounds of width $\sim 8t$. Away from half-filling the spectral weight transfers particularly from the lower band to the in-gap states at $\omega-\mu\sim-t$. The chemical potential lies in the upper quasiparticle band.}
\label{dos}
\end{figure}
Figure \ref{dos} shows the doping dependence of the DOS.
At half-filling $n=1.00^+$ (with appropriate shift of the chemical potential to the bottom of the upper band), the multiple scattering by spin fluctuations redistributes the MF-quasiparticle weight into the incoherent background in an energy range $\sim8t$ accompanied by sharp quasiparticle peaks at the edges of the insulating gap.
Based on the argument similar to Kane {\it et al.} for the $t$-$J$ model \cite{Kane89}, we can conclude that the quasiparticle residue and its width are the order of $t/U$ and $4t^2/U$, respectively.
The intensity of the incoherent background is considerably enhanced at $\omega-\mu\sim -7t$ and $3t$, which results from the coupling of the quasiparticle to interband particle-hole excitations across the gap \cite{Altmann95,Preuss95}.

With increasing electron doping, the position of the lower band moves down to lower energy with considerable reduction of the spectral weight.
The reduction of the spectral weight, particularly in the lower band, leads to a transfer into the insulating gap giving rise to the in-gap states at $\omega-\mu\sim-t$.
As the doping increases further the in-gap states gain more spectral weight.
Note that the chemical potential lies in the upper quasiparticle band.

\begin{figure}[t]
\includegraphics[width=8.5cm]{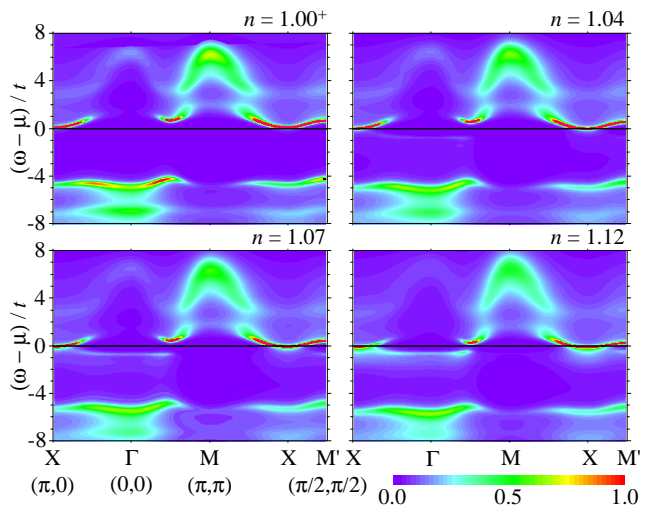}
\caption{The contour plot for the spectral intensity $A({\bm k},\omega)$ along the high-symmetry lines of the Brillouin zone. The quasiparticle band appears at the bottom (top) of the upper (lower) band for $n=1.00^+$. The AF shadow bands with rather weak intensity appear along X-$\Gamma$-M$'$ (M$'$-M-X) in the upper (lower) band.
Upon doping, the in-gap states at $\omega-\mu\sim-t$ together with the upper quasiparticle band acquire the features of a renormalized quasiparticle band similar to that in the paramagnetic phase.}
\label{akw}
\end{figure}
Figure \ref{akw} shows the contour plot for the spectral intensity $A({\bm k},\omega)$ along the high-symmetry lines of the Brillouin zone (BZ).
At $T=0$, the relevant scattering process creates spin excitations, which disturb of the AF background.
Since a coherent motion of a particle in the disturbed background feels an effectively smaller ordered moment, its excitation energy is smaller than that of the MF-states.
Hence we expect well-defined quasiparticle states at the edges of the renormalized gap unless the long-range order itself is broken.
For states with larger excitation energy, multiple scattering creating spin excitations dominates, and broad spectrum of incoherent states appears at energies far from the chemical potential.
For $n=1.00^+$ the spectrum that emerges has sharp quasiparticle bands at the edges of broad multiple spin excitation backgrounds as shown in Fig.~\ref{akw}.
The lower quasiparticle band, which would correspond to the charge-transfer band in the cuprates, has its highest energy and strongest intensity at the M$'$-points.
These features are confirmed by the ARPES experiment \cite{Armitage02}.
The qualitative form of weight redistribution in $A({\bm k},\omega)$ is consistent with the results of extensive quantum Monte Carlo simulation for the $t$-$U$ Hubbard model \cite{Bulut94a,Preuss95}.

\begin{figure}[t]
\includegraphics[width=8.5cm]{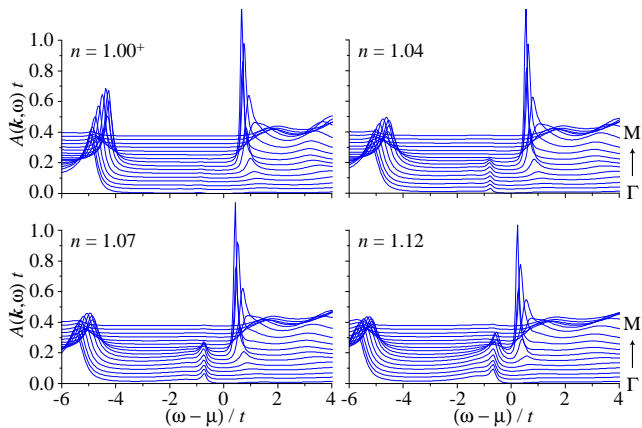}
\caption{The spectral intensity along $\Gamma(0,0)$-M$(\pi,\pi)$ line. As the doping increases, the in-gap states develop and become dispersive.
The in-gap states and the upper quasiparticle band exhibit features of a renormalized quasiparticle band similar to that in the paramagnetic phase.}
\label{akw2}
\end{figure}

Upon doping the chemical potential moves higher in the upper quasiparticle band, and small Fermi surface pockets appear close to the X-points as also found in the strong coupling calculations of Ref.~\cite{Lee03}.
These small pockets provide new decay channels through the creation of the particle-hole spin excitations in the upper quasiparticle band.
Then, a redistribution of the spectral weight again takes place particularly in the lower band, leading to a considerable reduction as shown in Figs.~\ref{dos} and \ref{akw}, which must be compensated by the appearance of spectral weight elsewhere, {\it i.e.}, in the in-gap state that appears at $\omega-\mu\sim -t$.
With further doping, the in-gap states become more visible in wide regions of the BZ, while the shadow bands become less visible except close to the chemical potential.
As a result, the in-gap states together with the upper quasiparticle band exhibit clear resemblance to a renormalized quasiparticle band in the paramagnetic phase with the pseudogap due to strong AF fluctuations (so-called hot spots) \cite{Armitage01,
%Moriya90,
Kampf90,Monthoux93,Kontani99,Moriya00}.
This is seen more clearly in Fig.~\ref{akw2}, in which the in-gap states develop and become dispersive with doping.

\begin{figure}
\includegraphics[width=8.5cm]{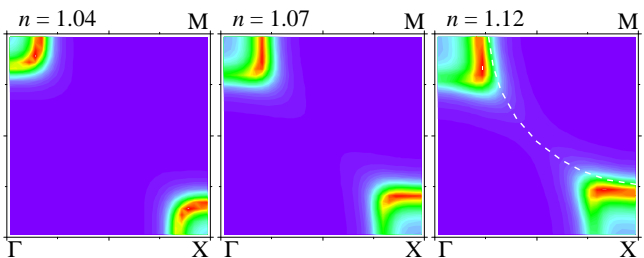}
\caption{The contour plot of $A({\bm k},0)$ at the chemical potential corresponding to a Fermi-surface plot. Upon doping the small Fermi surface pockets appear first around the X-points.
As the doping increases these pockets deform and become pieces of the paramagnetic-like Fermi surface indicated by the dashed line for $n=1.12$.}
\label{fs}
\end{figure}
Figure \ref{fs} shows the contour plot of $A({\bm k},0)$ at the chemical potential.
The shape of the strongest intensity corresponds to Fermi surfaces.
Upon doping the small Fermi surface pockets appear first around the X-points.
As the doping increases these Fermi surface pockets deform and become pieces of the paramagnetic-like Fermi surface.
The dashed line in Fig~\ref{fs} for $n=1.12$ denotes the Fermi surface of the non-interacting tight-binding dispersion in the paramagnetic phase.
The form of the Fermi surface deformation with doping agrees qualitatively with the ARPES measurement \cite{Armitage02}.

In order to elucidate the origin and the nature of the in-gap state, we consider a simple variational wavefunction \cite{Weng90}.
The MF ground state is expressed by filling up the states below the chemical potential,
\begin{equation}
|0\rangle=\prod_{{\bm k}}^{\rm MBZ}\prod_m^\pm\prod_\sigma^{\uparrow\downarrow} \theta(\mu_0-E_{{\bm k}m})\gamma_{{\bm k}\sigma m}^\dagger|{\rm vac}\rangle,
\end{equation}
where the MF-quasiparticle $\gamma$-operator is related to the $c$-operator by $c_{{\bm k}\sigma}=\sum_m^\pm v_{{\bm k}m}\gamma_{{\bm k}\sigma m}$ with $v_{{\bm k}m}=\sqrt{(1-m\xi^-_{{\bm k}}/E^0_{{\bm k}})/2}$.
The quasihole MF state in the lower band is described by $|{\bm p}\sigma-\rangle=\gamma_{{\bm p}\sigma-}|0\rangle$, which couples with multiple spin exchange states via the residual Hubbard interaction, $H_{\rm int}$.

Consider the variational wavefunction defined by
\begin{equation}
|\Psi_{{\bm p}\sigma-}\rangle=\sin\theta_{\bm p} |{\bm p}\sigma-\rangle - \cos\theta_{\bm p} |\phi_{{\bm p}\sigma}\rangle,
\end{equation}
where $|\phi_{{\bm p}\sigma}\rangle$ contains multiple spin excitations and $\theta_{\bm p}$ is a variational parameter.
A basic component of the $|\phi_{{\bm p}\sigma}\rangle$ is then given by applying $H_{\rm int}$ on $|{\bm p}\sigma-\rangle$, which generates states with a particle-hole spin excitation.
Among these states, the state involving a particle-hole spin excitation within the upper band makes the dominant contribution which lowers the energy of the $|\Psi_{{\bm p}\sigma-}\rangle$, {\it i.e.},
\begin{equation}
|\phi_{{\bm p}\sigma}\rangle=\frac{1}{\sqrt{A_{\bm p}}N}\sum_{{\bm k},{\bm q}}^{\rm MBZ}f({\bm p},{\bm k};{\bm q})\gamma_{{\bm p}+{\bm q}\sigma+}\gamma_{{\bm k}-{\bm q}\bar{\sigma}+}\gamma^\dagger_{{\bm k}\bar{\sigma}+}|0\rangle,
\end{equation}
where
\begin{equation}
f({\bm p},{\bm k};{\bm q})=p_{-}({\bm p}+{\bm q},{\bm p})p_{+}({\bm k}-{\bm q},{\bm k})-[{\bm q}\to{\bm q}+{\bm Q}],
\end{equation}
with the coherent factor $p_{\pm}({\bm k},{\bm k}')=v_{{\bm k}+}v_{{\bm k}'\pm}\pm v_{{\bm k}-}v_{{\bm k}'\mp}$.
Here $A_{\bm p}$ is the appropriate normalization factor.

\begin{figure}[t]
\includegraphics[width=8.5cm]{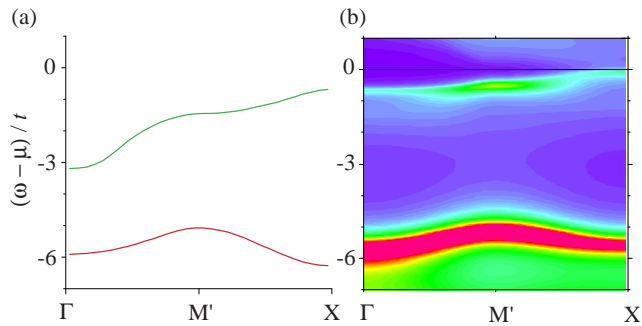}
\caption{The comparison of the quasihole dispersions for $n=1.12$: (a) evaluation by the variational wavefunction (green) and the MF energy (red), and (b) results of the SCBA calculation in the quasiparticle $\gamma$-representation.}
\label{qpdisp}
\end{figure}
Minimizing the total energy $E^{\rm tot}_{{\bm p}-}=\langle\Psi_{{\bm p}\sigma-}|H|\Psi_{{\bm p}\sigma-}\rangle$ with respect to $\theta_{\bm p}$, we obtain the renormalized quasihole energy, $\tilde{E}_{{\bm p}-}=-(E^{\rm tot}_{{\bm p}-}-E_0)$, where $E_0=\langle0|H|0\rangle$.
For small doping two holes in the $|\phi_{{\bm p}\sigma}\rangle$ have a momentum close to the X-points, i.e., ${\bm p}+{\bm q}\sim{\bm k}-{\bm q}\sim {\rm X}$.
Due to the momentum conservation the momentum of the particle is ${\bm k}\sim {\bm p}$.
Since the kinetic energy of the $|\phi_{{\bm p}\sigma}\rangle$ is the order of $E_{{\bm p}+}-\mu_0$ and the average of the Coulomb interaction is negligible, the average of the total energy of the $|\phi_{{\bm p}\sigma}\rangle$ is the order of $E_{{\bm p}+}-\mu_0$, which is the order of $U$ smaller than that of the hole in the lower band, $|{\bm p}\sigma-\rangle$.
Therefore, the most of the weight in the $|\Psi_{{\bm p}\sigma-}\rangle$ comes from the $|\phi_{{\bm p}\sigma}\rangle$ and we obtain an energy $\tilde{E}_{{\bm p}-}\sim -(E_{{\bm p}+}-\mu_0)$.

In Fig.~\ref{qpdisp} we compare the quasihole energies for $n=1.12$ (a) evaluated by the variational wavefunction with (b) the results of the SCBA calculation in the quasiparticle $\gamma$-representation.
The green (red) line in Fig.~\ref{qpdisp}(a) represents the renormalized (MF) quasihole band.
The dispersion of the renormalized quasihole band in Fig.~\ref{qpdisp}(a) follows the curve $-E_{{\bm p}+}$.
The variational evaluation of the quasihole energy agrees reasonably closely with the SCBA results except for region close to the $\Gamma$-point.
Since the states near the $\Gamma$-point are far from $\mu_0$, the simple variational wavefunction would be poor.
Note that a similar variational argument could be hold for hole dopings, where small hole pockets around the zone diagonals are involved instead (see Fig.~\ref{akw}).

Since the renormalization of the order parameter is neglected in this work, the AF phase survives up to a larger doping than the actual value ($n_{\rm c} \sim 1.12$).
However in reality, there exist well developed short-range order and strong AF fluctuations.
As the ARPES experiment is insensitive against the difference between a short-range order and true long-range order, our treatment with an unrenormalized MF order parameter mimics the presence of short-range order.
Nevertheless, the renormalization could be important for the smooth evolution of the Fermi surface near $n_{\rm c}$ accompanied by the appearance of spectral weight around the zone diagonals \cite{Armitage02,Damascelli03}.
These features are reproduced by MF approaches with renormalization of the ordered moment\cite{Kusko02}.
A more sophisticated theoretical treatment is needed to describe this smooth evolution of the spectra as the AF order changes rapidly from long to short range.

In summary, we have studied systematically the evolution of the single-particle spectrum in the doped antiferromagnetic insulator.
Starting from the MF antiferromagnetic insulator, we take into account multiple exchange of transverse spin excitations in the self energy.
Away from half-filling we observe small Fermi surface pockets near the X-points and simultaneously the in-gap quasiparticle states, which eventually evolve into a renormalized quasiparticle band similar to that in the paramagnetic phase.
We have proposed the simple variational wavefunction for the in-gap state, where the lower quasihole state admixes strongly with the upper quasihole state accompanied by a particle-hole excitation in the upper quasiparticle band.
We have demonstrated that the weak-coupling treatment starting from the AF insulator gives a good description of the single-particle spectrum of the electron-doped cuprates.

%.Acknowledgement
We would like to thank W. Brenig, Z.-X. Shen, K. Ueda, M. Sigrist and H. Monien for fruitful discussions.
This work is supported by the MaNEP program of the Swiss National Fund.


\begin{references}
\bibitem{Armitage02} N.P. Armitage {\it et al}., Phys. Rev. Lett. {\bf 88} 257001 (2002).
\bibitem{Damascelli03} A. Damascelli, Z. Hussain and Z.-X. Shen, Rev. Mod. Phys. {\bf 75} 473 (2003).
\bibitem{Honerkamp01} C. Honerkamp, Eur. Phys. J. B {\bf 21} 81 (2001).
\bibitem{Manske00} D. Manske, I. Eremin and K.H. Bennemann, Phys. Rev. B {\bf 62} 13922 (2000).
\bibitem{Kobayashi02} A. Kobayashi {\it et al}., J. Phys. Soc. Jpn. {\bf 71} 1640 (2002).
\bibitem{Chubukov95} A.V. Chubukov and K.A. Musaelian, J. Phys. Condens. Matter {\bf 7} 133 (1995).
\bibitem{Furukawa98} N. Furukawa, T.M. Rice and M. Salmhofer, Phys. Rev. Lett. {\bf 81} 3195 (1998).
\bibitem{Honerkamp01a} C. Honerkamp {\it et al}., Phys. Rev. B {\bf 63} 35109 (2001).
\bibitem{Yanase01} Y. Yanase and K. Yamada, J. Phys. Soc. Jpn. {\bf 70} 1659 (2001).
\bibitem{Kobayashi01} A. Kobayashi {\it et al}., J. Phys. Soc. Jpn. {\bf 70} 1214 (2001).
\bibitem{Lee03} T.K. Lee, C.-M. Ho and N. Nagaosa {\bf 90} 067001 (2003).
\bibitem{Kusko02} C. Kusko {\it et al}., Phys. Rev. B {\bf 66} 140513 (2002).
\bibitem{Schrieffer89} J.R. Schrieffer, X.G. Wen and S.C. Zhang, Phys. Rev. B {\bf 39} 11663 (1989).
\bibitem{Bulut94} N. Bulut, D.J. Scalapino and S.R. White, Phys. Rev. Lett. {\bf 73} 748 (1994).
\bibitem{Altmann95} J. Altmann {\it et al}., Phys. Rev. B {\bf 52} 7395 (1995).
\bibitem{Brenig95} W. Brenig, Phys. Rep. {\bf 251} 153 (1995).
\bibitem{Bulut93} N. Bulut, D.J. Scalapino and S.R. White, Phys. Rev. B {\bf 47} 2742 (1993).
\bibitem{Kane89} C.L. Kane, P.A. Lee and N. Read, Phys. Rev. B {\bf 39} 6880 (1989).
\bibitem{Preuss95} R. Preuss, W. Hanke and W. von der Linden, Phys. Rev. Lett. {\bf 75} 1344 (1995).
\bibitem{Bulut94a} N. Bulut, D.J. Scalapino and S.R. White, Phys. Rev. B {\bf 50} 7215 (1994).
\bibitem{Armitage01} N.P. Armitage {\it et al}., Phys. Rev. Lett. {\bf 87} 147003 (2001).
%\bibitem{Moriya90} T. Moriya, Y. Takahashi and K. Ueda, J. Phys. Soc. Jpn. {\bf 59} 2905 (1990).
\bibitem{Monthoux93} P. Monthoux and D. Pines, Phys. Rev. B {\bf 47} 6069 (1993).
\bibitem{Kontani99} H. Kontani, K. Kanki and K. Ueda, Phys. Rev. B {\bf 59} 14723 (1999).
\bibitem{Moriya00} T. Moriya and K. Ueda, Adv. Phys. {\bf 49} 555 (2000).
\bibitem{Kampf90} A.P. Kampf and J.R. Schrieffer, Phys. Rev. B {\bf 42} 7967 (1990).
\bibitem{Weng90} Z.Y. Weng, C.S. Ting and T.K. Lee, Phys. Rev. B {\bf 41} 1990 (1990).
\end{references}
\end{document}